\documentclass[final,1p,times]{elsarticle} 
\usepackage{graphicx} 
\usepackage{amssymb} 
\usepackage{amsthm} 
\usepackage{lineno} 

\journal{Nuclear Physics A} 
\begin{document} 

\begin{frontmatter} 


\title{High-$P_T$ Physics with Identified Particles}

\author{Rainer J.\ Fries$^{a,b}$ and Wei Liu$^a$}

\address[a]{Cyclotron Institute, Texas A\&M University, College Station, TX 77843,
USA}
\address[b]{RIKEN/BNL Research Center, Brookhaven National Laboratory, Upton, NY
11973, USA}

\begin{abstract} 
The suppression of high-$P_T$ particles in heavy ion collisions was one of 
the key discoveries at the Relativistic Heavy Ion Collider. This is usually 
parameterized by the average rate of momentum-transfer squared  
to this particle, $\hat q$. 
Here we argue that measurements of identified particles at high $P_T$ can lead 
to complementary information about the medium. The leading particle of a 
jet can change its identity through interactions with the medium. 
Tracing such flavor conversions could allow us to constrain
the mean free path.
Here we review the basic concepts of flavor conversions and discuss 
applications to particle ratios and elliptic flow. We make a
prediction that strangeness is enhanced at high $P_T$ at RHIC 
energies while its elliptic flow is suppressed.
\end{abstract} 

\end{frontmatter} 


For the past decade high momentum particles and jets have been
used to probe the quark gluon plasma (QGP) phase created at the
Relativistic Heavy Ion Collider (RHIC). The energy loss of a fast
parton suffered in the medium carries information about the
typical momentum transfer $\mu$ along the path, more precisely about
the transport coefficient $\hat q = \mu^2/\lambda$ \cite{Wang:1991xy,BDMPS:96,Zakharov:96,Wiedemann:2000tf,gyulassy,AMY:02}. We have recently
argued that the mean free path $\lambda$ of a fast parton 
could be determined separately by measuring the change in hadro-chemistry 
induced by the medium \cite{weiliu,fries:09}.

Here we discuss a model based on conversions of the leading particle
of a jet. Just as partons can lose energy through collisions and induced
radiation, they can scatter through channels in which the identity of 
the fastest parton in the initial and final state are not the same.
Examples are binary collisions like $q+\bar q \leftrightarrow g+g$ or
$q+g \leftrightarrow g+q$ which can lead to conversions of quarks into 
gluons and vice versa. Here the first parton on each side has a large 
momentum (the leading jet parton) and the second parton in the initial
state is a thermal parton from the quark gluon plasma.
The rate of flavor conversions depends on the mean free
path $\lambda$ of fast partons.

\begin{figure}[t]
\centerline{
\includegraphics[width=5.0cm,angle=-90]{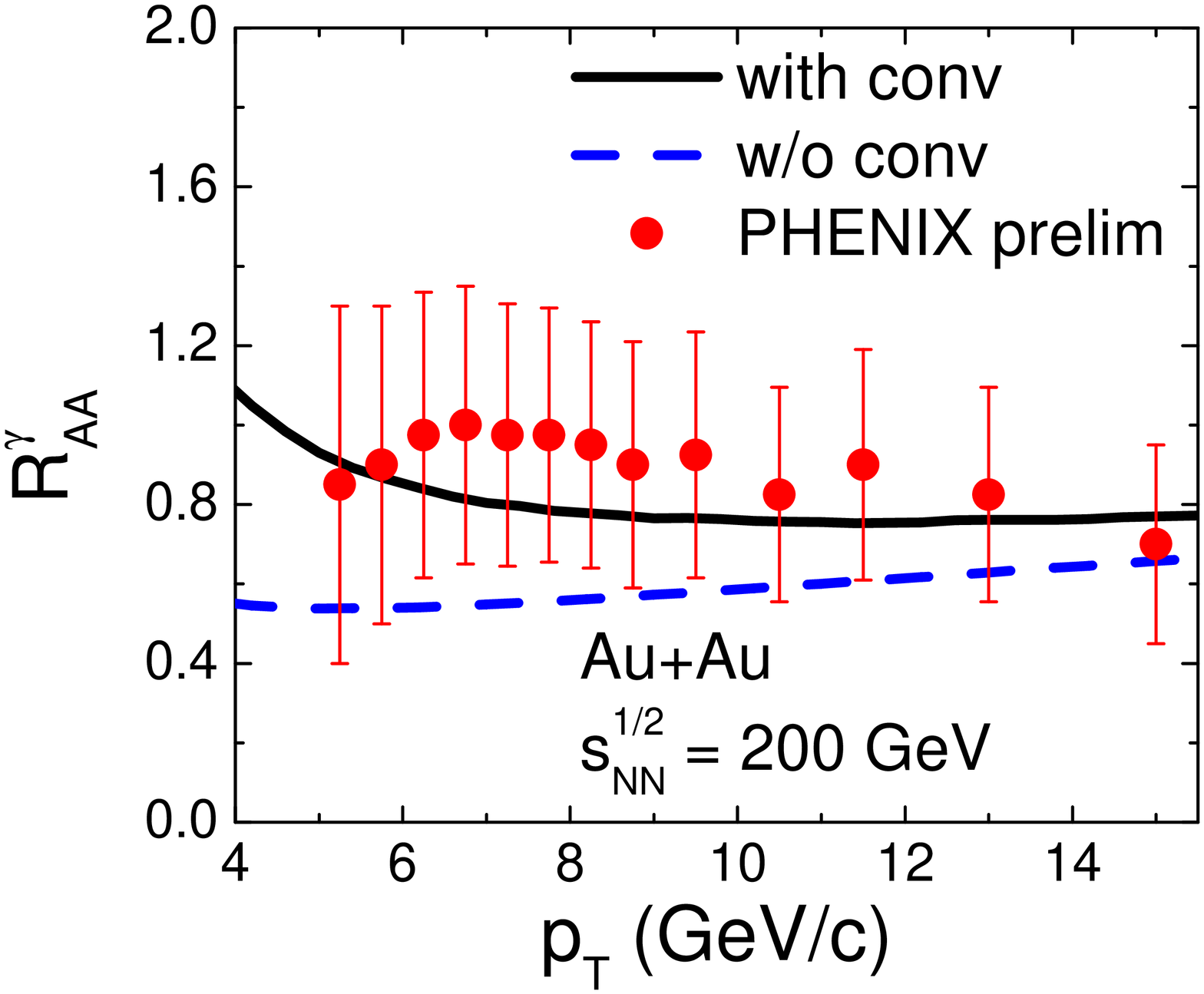}
\includegraphics[width=5.0cm,angle=-90]{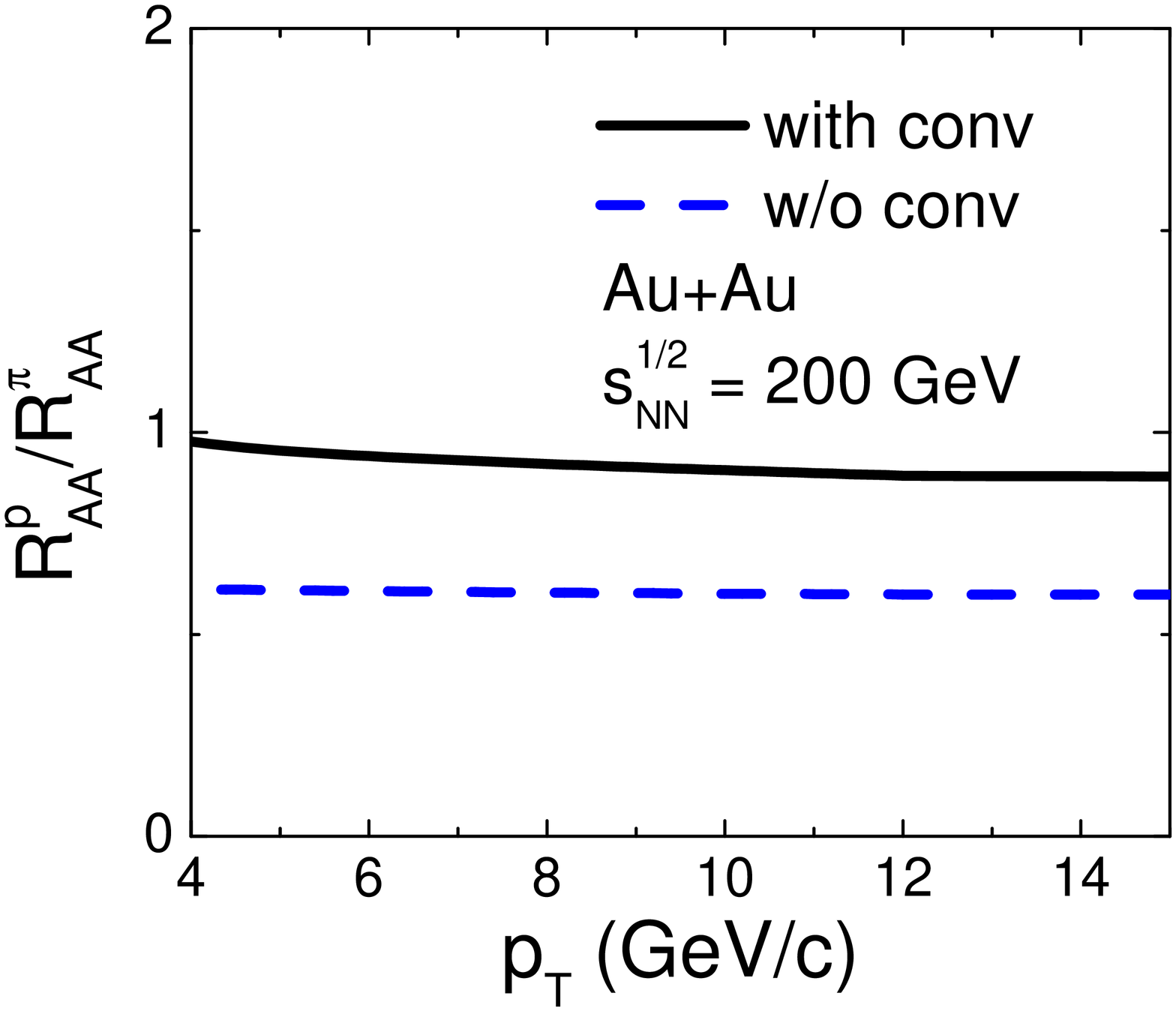}}
\caption{Left panel: The nuclear modification factor $R_{AA}$ for direct
photons with and without conversions switched on, calculated in the model
introduced in \cite{weiliu} (preliminary PHENIX data from \cite{Adler:2005ig}).
Right panel: The ratio of nuclear modification
factors for protons and pions is approaching one if conversions are allowed.}
\label{fig:1}
\end{figure}

Conversions between quarks and gluons should obscure their different
color factors coupling them to the medium. Instead of a relative factor
9/4 in $\hat q$ only an average color factor should be observable in
a long enough medium. It was pointed out that a larger quenching for gluons
could be reflected in more suppression of protons compared to pions
given the preference of gluon to proton fragmentation in modern
fragmentation functions \cite{akk}. Jet conversions should soften 
this effect and increase the proton to pion ration in central
collisions \cite{weiliu1,cmk}. Fig.\ \ref{fig:1} shows the ratio of 
nuclear modification factors $R_{AA}$ for protons and pions with and 
without conversions between quarks and gluons taken into account 
\cite{weiliu}. Clearly, flavor conversions lead to less proton 
suppression.

Conversions had also been discussed before for photons and dileptons.
Fast quarks and gluons can create real or virtual photons through Compton
and annihilation processes with the medium, $q+g \to \gamma+q$, $q+\bar q \to
\gamma+g$. It has been realized
over the years that this process can make a large contribution
to the overall yield of photons or dileptons \cite{fries1,FMS:05,SGF:02,
GAFS:04,simon,Turbide:2007mi}.
Fig.\ \ref{fig:1} also shows the nuclear modification factor for direct photons
with and without jet conversions into photons from the computation described
in \cite{weiliu} together with data from PHENIX.

\begin{figure}[t]
\centerline{
\includegraphics[width=7.0cm]{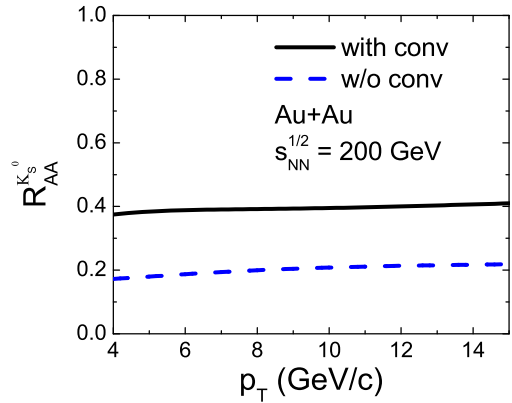}}
\caption{$R_{AA}$ for neutral kaons with and without conversion processes
allowed. The strangeness in the jet sample is driven towards equilibrium
by coupling it chemically to the quark gluon plasma.}
\label{fig:2}
\end{figure}

Most recently we predicted an enhancement of strange hadrons from jet
conversions at RHIC energies. This is driven by a strong chemical gradient
between the jet sample at RHIC and the quark gluon plasma. While strangeness
is equilibrated in the latter, it contributes only a small fraction ($< 
5\%$) to the former. Elastic channels like $g+g \to s+\bar s$ and in
particular kick-out reactions like $g+s \to s+g$ can lead to an enrichment
of strangeness in the jet sample. Obviously, for a sufficiently long medium
the jet sample would be driven toward chemical equilibrium through
interactions with the chemically equilibrated medium.
Fig.\ \ref{fig:2} shows the expected nuclear modification factor for neutral
kaons at RHIC. For the perturbative elastic rates chosen in \cite{weiliu}
with a $K$ factor of 4 we see a clear enhancement extending above 
10 GeV/$c$.

We have also checked the effect of conversions on heavy quark production
\cite{Liu:2008bw}. We do not find any significant yields of charm or 
bottom quarks at high $P_T$ from these processes. The same is true even at
LHC energies. The reason is that thresholds and low center of mass 
energies suppress pair production in interactions of jets with the medium, 
and kick-out reactions suffer from small heavy quark densities in the
medium to begin with. Even for charm at LHC we do not expect chemical 
equilibration and there is no large chemical gradient between heavy 
quarks in jets and the medium.

\begin{figure}[t]
\centerline{
\includegraphics[width=5.5cm,angle=-90]{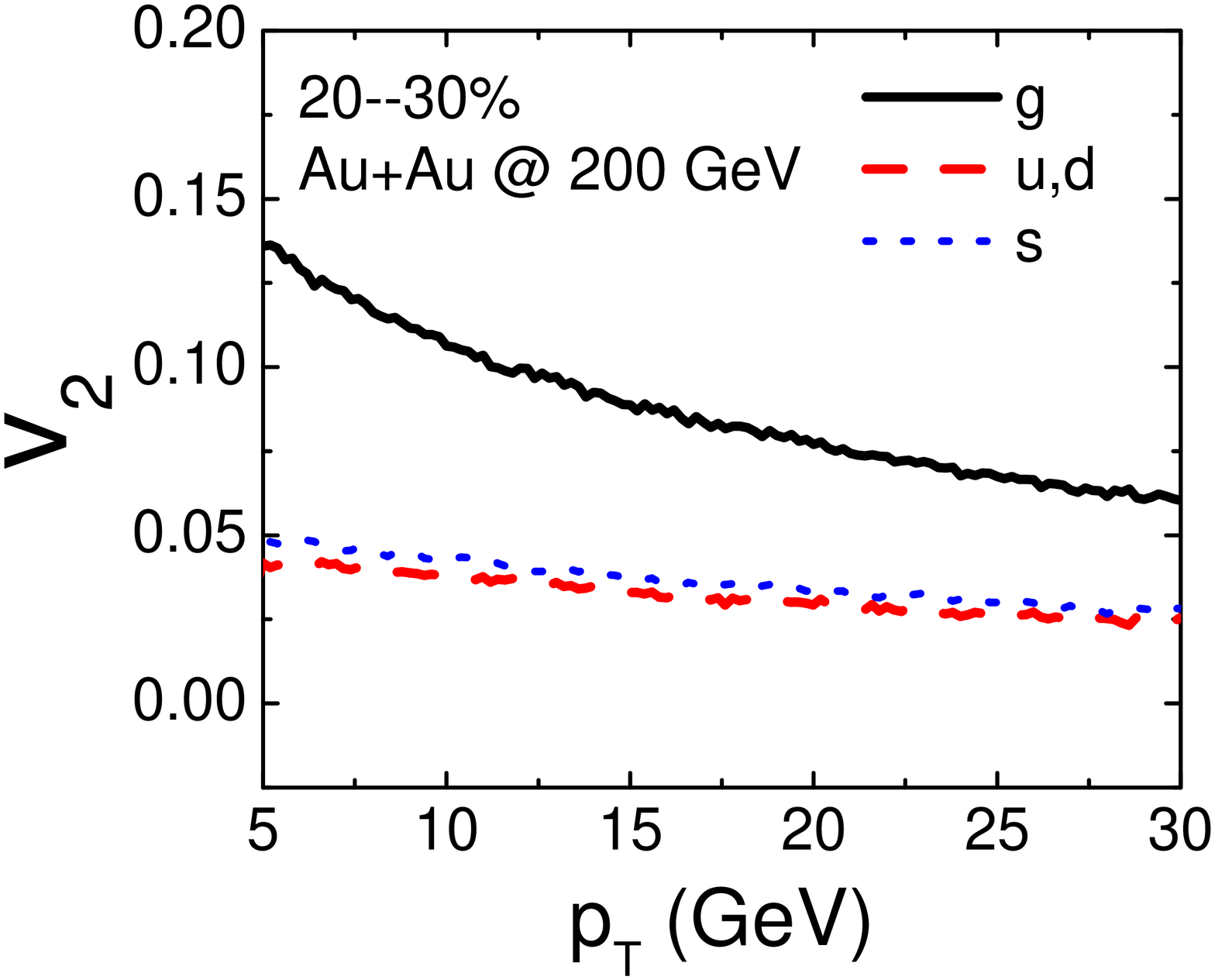}
\includegraphics[width=5.5cm,angle=-90]{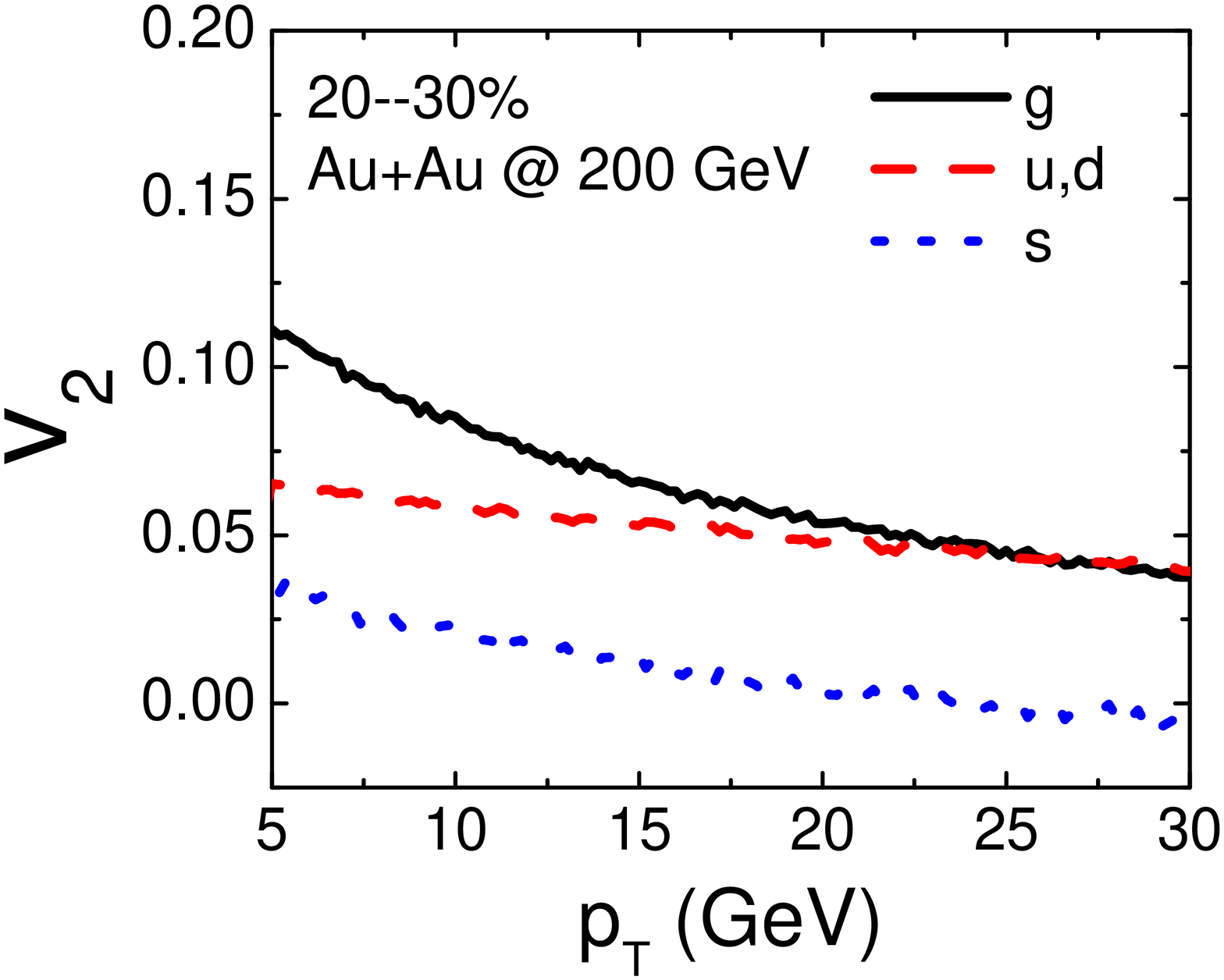}}
\caption{Left panel: The azimuthal asymmetry $v_2$ for light quarks, strange
quarks and gluons without conversions. Right panel: the same with conversions.
The $v_2$ for light quarks and gluons is not similar, while strange quarks
exhibit a suppression.}
\label{fig:3}
\end{figure}

\begin{figure}[t]
\centerline{
\includegraphics[width=6.0cm,angle=-90]{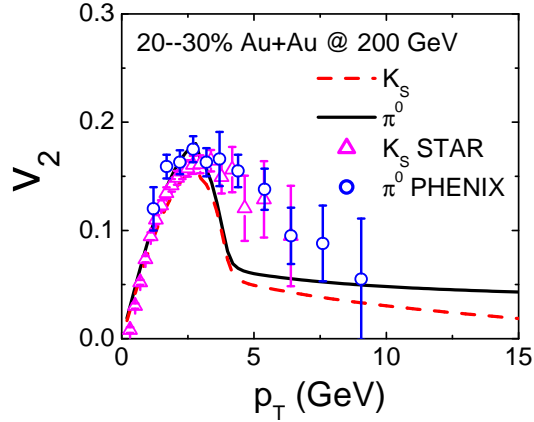}}
\caption{The resulting azimuthal asymmetry $v_2$ for kaons which is
expected to be suppressed compared to that of pions. Data from
\cite{david,abelev}.}
\label{fig:4}
\end{figure}

We have also studied the effect of flavor conversions on the azimuthal 
asymmetry $v_2$. It was first pointed out in Ref.\ \cite{Turbide:2005bz} 
that photons from jet-medium interactions should be more abundantly produced
in the direction where the medium is thicker, leading to a negative 
contribution to $v_2$. See \cite{Chatterjee:2005de} for comprehensive
calculations of photon elliptic flow.
It is important to realize that this mechanism 
is generally true for all particles produced in jet-medium interactions
\cite{Liu:2008kj}.
In particular, we expect it to hold for additional strange hadrons produced
at RHIC. Fig.\ \ref{fig:3} shows the effect of flavor conversions on the
azimuthal asymmetry $v_2$ of up, down and strange quarks as well as gluons.
As predicted, conversions make light quarks and gluons behave similarly
while there is a significant decrease in the $v_2$ of strange quarks due
to the additional large yield with a negative contribution. Note that
the total $v_2$ is determined from summing up all sources of particles 
both with positive and negative $v_2$. Fig.\ \ref{fig:4} shows the
resulting azimuthal asymmetry $v_2$ that we expect for kaons at RHIC
\cite{Liu:2008kj}.

In summary, we have shown that jet conversions can lead to measurable
signatures in hadron production at high $P_T$. We predict an enhancement
of strange hadrons at high $P_T$ at RHIC and a suppression of their
azimuthal asymmetry coefficient $v_2$.

\section*{Acknowledgments} 
We acknowledge support by the RIKEN/BNL Research Center and DOE grant 
DE-AC02-98CH10886.

\end{document}